\documentclass[natbib]{llncs}
\usepackage{amsmath}
\title{ATLAS: Interactive and Educational Linear Algebra System Containing Non-Standard Methods}
\titlerunning{ATLAS: Educational Linear Algebra}
\author{Akhilesh Pai\inst{1} \and James H. Davenport\inst{1}}
\institute{
${}^1$  University of Bath, Bath BA2 7AY, UK\\
  \email{abp43@bath.ac.uk}; \email{J.H.Davenport@bath.ac.uk} 
}
\begin{document}

\maketitle

\begin{abstract}
While there are numerous linear algebra teaching tools, they tend to be focused on the basics, and not handle the more advanced aspects. This project aims to fill that gap, focusing specifically on methods like Strassen's fast matrix multiplication.
\end{abstract}
\section{Introduction}
There are numerous linear algebra teaching tools, of which we make a brief survey of the more relevant ones.
\subsection{Maple/Tutor}
Maple, a major computer algebra system, has various linear algebra tools, but all of the ``$O(n^3)$ for multiplication'' family, despite evidence \cite{Tonksetal2017a} that, in an algebraic rather than numerical context, the Strassen--Winograd \cite{Strassen1969,Winograd1971} method can in practice be faster for reasonably small examples.

Maple also has a feature called Tutor, which allows users to view the step-by-step solutions to a problem. The Tutor feature is available for some of the linear algebra functions, such as calculation of eigenvalues \cite{eigentutor}.  This is a very useful feature because it allows users to directly view how a problem
is solved each step of the way, allowing them to learn the method that led to the solution instead of just
getting the correct answer. This feature may be responsible for the popularity of Maple amongst students,
as it can serve as a key tool in education. However, this feature is also limited to only standard methods.
Incorporating a Tutor-style feature in a linear algebra system, however, combined with the inclusion of
alternative methods could create a platform through which different methods can be compared adjacent
to each other. This will not only encourage users to learn how the solution is found, but also give users
the opportunity to learn entirely new methods that are, in some cases, more efficient.
\subsection{CoCalc}
Created by the developers of SageMath \cite{sagemath}, CoCalc \cite{Stein2018a} is a CAS that is unique amongst its competition because it is web-based and therefore does not need to be downloaded or installed \cite{cocalc}. This immediately highlights one of the advantages of CoCalc over competitors such as Maple and REDUCE, which is that it is accessible on any computer system with a stable internet connection without needing a powerful computer. Therefore, it far surpasses Maple and REDUCE in its portability. CoCalc's relation to SageMath is that it is the cloud version of SageMath \cite{sagebook}.

Another advantage of CoCalc over its competition is that CoCalc allows collaboration, so projects within CoCalc can be worked on by multiple collaborators simultaneously \cite{collab} and CoCalc will record the history of the file through its Time Travel feature \cite{timetravel}, which also shows which collaborators made which changes to a given file allowing the collaborators to view the full development of their project from start to finish and revisit an older version of the file if they need to. This feature expands the target audience from individual students and researchers to groups of students and researchers. Furthermore, its use in educational settings would allow teachers to create notebooks and conduct their lessons entirely on CoCalc by giving students the chance to interactive with and control the notebook. This is most notably achieved through the ability to create courses that students can be a part of, create handouts and assignments for students to complete which can then be graded either by the teacher or through the builtin peer grading system that randomly assigns the assignments to other students \cite{sagebook}. However, these features are part of the paid functionality. In this respect, CoCalc has many similarities to the e-learning platform, Moodle, which is used by many universities such as the University of Bath.

\subsection{Symbolic Math Toolbox (MATLAB)}
The Symbolic Math Toolbox (SMT) by MathWorks \cite{smt} is a toolkit that works in conjunction with the popular programming language and computing environment MATLAB \cite{matlabweb} and amongst a variety of features, it offers linear algebra functionality. This feature within SMT includes the ability to find the determinant, the inverse, multiply matrices, and calculate the eigenvalues and eigenvectors of a matrix, in addition to offering the capability to calculate the characteristic equation and solve linear equations which are in matrix form \cite{matlab_linear}. However, one important aspect to note is that this toolbox shares the same flaw as systems such as CoCalc and REDUCE, which is that the users would need to specifically learn a new programming language in order to utilise its range of capabilities. Also, SMT does not offer any formula editing features, therefore users would be left with no other choice than to learn and use the SMT code. While the code within SMT is not particularly complex, this does add the additional barrier of having to learn a new programming language. Furthermore, another problem with SMT is that it does not offer alternative methods, such as Strassen's method, limiting itself to only standard methods of matrix multiplication. A further drawback of SMT from our point of view is that it does not offer a step-by-step view of the solution for a given problem, so it only provides the output for any given input. This means that SMT, similar to REDUCE and CoCalc, acts as a problem solver which, while being a useful part of the learning process, does not serve to teach users about the methods and concepts being used.

\cite{matlab} conducted a study about the use of MATLAB to teach 23 university students linear algebra with the aim to understand the perception of students about whether a software like MATLAB could be used to supplement the learning in their course. In general, he found that his students enjoyed using MATLAB to supplement their learning. As part of the study, students were given 3 MATLAB-based tasks to complete: use of Gaussian elimination and inverse computing methods on a $20\times20$ matrix, solving systems of linear equations using Cramer's rule and using the SVD method for image compression. The students were also given examples of how small scale versions of those problems could be solved using MATLAB. The results for this study were primarily collected through surveys asking whether the students liked MATLAB, whether it helped them understand linear algebra and whether they like to use MATLAB for their homework. The results of the study show that students who usually earned lower grades preferred MATLAB to students who usually earned higher grades, as they found it helped them with solving larger problems and seeing the answer to a problem without having to work it out using pen and paper. A student also remarked that it helped them understand linear algebra without needing to understand the algebraic minutiae of the method involved. However, for the students who were displeased with MATLAB, some said that they did not like MATLAB because they are "not good at computers", but the most important feedback was that a student remarked that MATLAB simply gives the answer to a problem without any explanations of the procedure. This highlights, as previously stated, a fundamental flaw within MATLAB and many similar systems, which is that providing the answer to a problem is not sufficient for teaching students the method. In the context of this study, the highly achieving students do well because they usually understand the method that leads to the answer, whereas the low achieving students are happy to just get the answer right. This is further exemplified by a 'C' grade student commenting "Can we use this to do our final exam?", as they would be able to input the problem and get an immediate answer without concerning themselves with the method. So high achievers want an explanation provided to them so that they can understand how MATLAB went from the initial problem to the final solution. Also, for the students who are "not good at computers", it is important to realise that not all students who are using these programs will be familiar with computers or programming. So, in cases where the student is expected to learn an entirely new programming language that they have not seen before to utilise the functionality of MATLAB or any other system, it can be off putting for them.

\section{State of ATLAS}
As of paper submission, the ATLAS system supports the calculation of determinants, inverses, eigenvectors and eigenvalues, in addition to providing the capability of matrix multiplication and solving systems of linear equations. In the case of calculating determinants, Laplace expansion, Sarrus' method and LU decomposition are supported, allowing users to compare all 3 methods adjacent to each other and view each step of each solution simultaneously. This is also true of matrix multiplication, which supports dot product and Strassen's method, and calculation of inverses using both the Cramer's rule and the Cayley-Hamilton theorem. Whilst the program has not yet reached a state of completion, once complete, multiple methods will also be included for the remaining 3 functions to allow comparison of methods.

Currently, unit testing has been carried out on individual, isolated components of ATLAS to ensure that all inputs are handled correctly, including erroneous inputs and edge cases, with bug fixes applied where necessary. Neither comprehensive integration testing nor user testing has been carried out as of yet, but this will be carried out in greater depth as the program nears completion, with 10 users already gathered for user testing to be conducted.

There is an aspiration to extend the system beyond the current suite of capabilities to look at and compare methods that are considered numerically good or bad. The purpose of this extension would be to understand which methods are superior to others when dealing with floating point numbers within matrices. This is because, there are cases where certain methods can lead to drastically incorrect results when given specific inputs due to the manner in which the method operates. Therefore, by implementing this functionality, users could be given a greater insight into the effectiveness of different methods.

Any future plans for ATLAS would revolve around the idea of implementing more non-standard methods, especially those that are newly developed to ensure that users have access to the newest and most efficient (or numerically most stable) methods available. 

Additionally, there is a desire to improve the portability of ATLAS by creating a web-based equivalent that can be accessed from anywhere without the need to transfer the files of the program. The development of CoCalc out of SageMath illustrates this. This development would help to increase the reach of ATLAS not only because it would be accessible from anywhere, but also because less computational power would be needed to run the program. 
\section{Pedagogical Application}
Even the simplest algorithm with non-na\"\i{}ve complexity, the Strassen--Winograd multiplication algorithm, is hard to understand and motivate. This outline is how the second author does it.
\begin{enumerate}
\item Matrix multiplication $A\times B$ is linear in $B$.
\item The Strassen--Winograd are also linear in $B$ (slightly tedious, but purely a matter of local verification.
\item Hence it suffices to verify that Strassen--Winograd  computes
$$
\begin{pmatrix}
a&b\cr c&d
\end{pmatrix}\times
\begin{pmatrix}
1&0\cr0&0
\end{pmatrix}
=\begin{pmatrix}
a&0\cr c&0
\end{pmatrix}
$$
(and three analogues).
\item This is tedious by hand, but easy with this tool.
\end{enumerate}
\section{Demonstration}
A demonstration of this tool, the main way of seeing it, can be found at \hfil\break\url{https://drive.google.com/drive/folders/1lBbcWdRd7VwVb8il3HHZgZO8njOx64Pb}.

\section{Conclusion}
The development of ATLAS was started with the aim of aiding students with learning various different linear algebra methods through active examples, in addition to allowing people such as researchers to obtain a broader linear algebra toolkit to apply for more complex problems for which standard methods would prove to be too inefficient or laborious. ATLAS aims to address several problems identified during the literature survey, such as the inability to compare multiple methods simultaneously and, in many cases, produce step-by-step solutions for a problem. Whilst step-by-step solutions are available within the popular computer algebra system Maple, the functionality is limited to a small subset of methods. These problems have been successfully addressed within ATLAS to create a linear algebra system with a simple and easy to use interface.

\def\href{\url}\def\path{\url}
\bibliography{references}
\end{document}